\documentclass[conference, twocolumn]{IEEEtran}

\IEEEoverridecommandlockouts

\usepackage{cite}
\usepackage{amsmath,amssymb,amsfonts}
\usepackage{gensymb}
\usepackage{float}
\usepackage{stfloats}

\usepackage{array}
\usepackage{algorithmic}
\usepackage{graphicx}
\usepackage{textcomp}
\usepackage{xcolor}
\usepackage[hidelinks]{hyperref}
\usepackage{cancel}
\usepackage{multirow}

\usepackage{amsthm}
\theoremstyle{definition}

\ifCLASSOPTIONcompsoc
    \usepackage[caption=false, font=normalsize, labelfont=sf, textfont=sf]{subfig}
\else
\usepackage[caption=false, font=footnotesize]{subfig}
\fi

\DeclareMathOperator{\sinc}{sinc}

\DeclareMathOperator{\Si}{Si}

\begin{document}

\title{Approximation of the Narrowband Velocity Ambiguity Function in Near-Field Sensing Systems}

\author{
    \IEEEauthorblockN{Marcin Wachowiak\IEEEauthorrefmark{1}\IEEEauthorrefmark{2}, 
    André Bourdoux\IEEEauthorrefmark{1}, 
    Sofie Pollin\IEEEauthorrefmark{2}\IEEEauthorrefmark{1}}
    \IEEEauthorblockN{
        \IEEEauthorrefmark{1} imec, Kapeldreef 75, 
        3001 Leuven, Belgium \\
        \IEEEauthorrefmark{2} Department of Electrical Engineering, 
        KU Leuven, Belgium \\
        Email: marcin.wachowiak@imec.be 
        }   
}

\maketitle

\begin{abstract}
This work investigates the velocity ambiguity function in the radiative near-field of a uniform linear antenna array. 
First, the general ambiguity function is formulated and the velocity cross-section at a target position is considered. 
The velocity vector is decomposed into radial and tangential components, which are analyzed separately. 
Next, the closed-form approximations for the velocity ambiguity function are derived and shown to be accurate despite relying on far-field approximations. 
The results show that the near-field regime does not improve radial velocity estimation, but enables estimation of the tangential velocity. 
Finally, the velocity ambiguity functions and the achievable resolution are analyzed for different array configurations. 
For tangential velocity estimation, MIMO and a collection of SISO radars offer improved ambiguity function and resolution as compared to the SIMO/MISO configuration.
\end{abstract}


\section{Introduction}

\subsection{Problem Statement}

To achieve higher resolution and better sensing performance, modern sensing systems employ an increasingly large number of antenna elements \cite{rec_adv_4d_mimo_rad}. 
Consequently, these systems are becoming increasingly distributed and the total physical aperture of the system increases.
As the total aperture increases, the radiative near-field region of the antenna array expands \cite{primer_on_nf_bf} and may encompass targets of interest.
In the radiative near-field, the propagation is modeled by a spherical wave, resulting in nonlinear phase difference across the antenna elements \cite{primer_on_nf_bf, nf_sensing_w_bw}. The nonlinear phase difference produces a range-dependent array factor, creating new opportunities for sensing and communication systems \cite{nf_localization_and_sensing_sp_mag, nf_multiplexing}.
In particular, the range-dependent array factor enables discrimination of targets in range without relying on bandwidth \cite{mw_nf_mimo_resolution}.
The benefits of the range-dependent array factor on the sensing ambiguity function and range resolution have already been studied and quantified \cite{nf_sensing_w_bw}.

The effects of the near-field operation on the Doppler and velocity ambiguity function remain relatively unexplored. To complete the picture, the Doppler/velocity ambiguity in the near-field regime should be investigated. The potential benefits of the near-field operation on the velocity ambiguity function and achievable resolution should be quantified and discussed.

\subsection{Relevant works}

The performance bounds of the near-field velocity sensing in terms of the Cramér–Rao lower bound (CRLB) are provided in \cite{nf_dopp_perf_bounds}. 
The CRLB is also used in \cite{bistatic_dopp_elaa}, to analyze the Doppler estimation accuracy of a bistatic system with a single large aperture.
A maximum likelihood-based method for near-field velocity sensing is proposed in \cite{nf_dopp_pred_beamform}, where the velocity information is utilized for predictive beamforming.
Near-field beamforming with three-dimensional velocity sensing is considered in  \cite{nf_bf_3d_vel_sensing}, with the application to unmanned aerial vehicle (UAV) localization and communications.
A subarray-based method to estimate the localization and velocity of a target is proposed in \cite{nf_joint_loc_and_vel_est_in_xlmimo}. To reduce the computational complexity, it employs subarray processing, which is based on a far-field channel model.
Joint localization and velocity estimation of a signal source in the near-field of the large antenna array is discussed in \cite{nf_source_localization_and_vel_est_elaa}.
Near-field velocity sensing is further exploited in \cite{nf_vel_sens_and_dopp_aware_nf_localization} to improve the localization accuracy.
Doppler-only extended target tracking is investigated in \cite{dopp_only_extended_target_tracking}, demonstrating the feasibility of target localization estimation using only the near-field Doppler information.

Many of the listed works analyze the CRLB for radial and transverse velocities and propose advanced algorithms to exploit the additional information provided by the velocity vectors. However, they do not discuss the velocity ambiguity function or the achievable velocity resolution. Therefore, it is essential to analyze the velocity ambiguity function in the near-field to assess the sidelobes and quantify the resolution, to determine the multi-target sensing performance. 

\subsection{Contributions}

This work investigates the velocity ambiguity function in the near-field of a ULA.
First, the velocity ambiguity function is formulated for a distributed narrowband system.
Next, the radial and tangential velocity cross-sections of the ambiguity function are analyzed and approximated with closed-form expressions. The results show that the radial velocity is largely unaffected by the near-field regime, while the transverse velocity depends on the target distance and array configuration.
The performance of several array configurations such as SIMO/MISO, MIMO and a collection of separated SISO radars is compared.
The velocity ambiguity function and resolution are derived and compared across array configurations.
The approximations of the ambiguity function and resolution are shown to remain accurate over a wide range of system and target parameters.

\section{Signal model}

\subsection{Near-field geometry}
\label{sec:sig_mod}
Consider a set of $M$ transmit (TX) and $N$ receive (RX) antennas denoted as $\mathcal{M}$ and $\mathcal{N}$, which form a large transmit and receive aperture. The antennas are arranged in a uniform linear array (ULA) along the X axis. The position of the antennas is given by $\mathbf{p}_m^{\mathrm{tx}} = \left[ m d_{\mathrm{a}}, 0\right]^T$ and $\mathbf{p}_n^{\mathrm{rx}} = \left[ n d_{\mathrm{a}}, 0\right]^T$, where $d_{\mathrm{a}}$ is the element spacing and $m$ and $n$ are the antenna indices. 
The following derivations are performed with respect to the transmit aperture and index $m$. However, they can be easily generalized to the receive aperture by relabeling the index as $n$.
The antenna indexing is symmetric, with the $0$th antenna located at the aperture geometric center.
Consider some point of interest expressed in polar coordinates by $\mathbf{p} =\left[d, \theta \right]^T$, with the azimuth angle $\theta$ defined with respect to the array normal.
Given the antenna array geometry, the distance to point $\mathbf{p}$ from the $m$th antenna can be written as
\begin{align}
    \label{eq:dist_full}
    d_m^{\mathrm{tx}} = \sqrt{d^2 + (m d_{\mathrm{a}})^2 - 2 d(m d_{\mathrm{a}}) \sin{(\theta)}}.
\end{align}
The bistatic distance is denoted as $d_{m, n}(\mathbf{p}) = d_m^{\mathrm{tx}} + d_n^{\mathrm{rx}}$.

The point of interest $\mathbf{p}$ is located within the radiative near-field of the array. In the near-field of the array, due to the extent of the aperture and proximity of the point of interest, the propagation is modeled using a spherical wave model.
Typically, the radiative near-field region of the array starts at the minimum distance at which gain variations across the aperture due to distance are negligible, which is $1.2D$ \cite{fresnel_lower_bound, primer_on_nf_bf}, where $D$ is the total array aperture. The upper bound of the near-field region is given by the maximum distance at which the beamdepth remains finite, which for a ULA is $d_{\mathrm{FA}} / 6.952$, \cite{mw_nf_mimo_resolution} where $d_{\mathrm{FA}}$ is the Fraunhofer distance given by $2D^2 / \lambda$ and $\lambda$ is the wavelength.
Within the near-field region the distance from \eqref{eq:dist_full} can be approximated using a second-order (quadratic) Taylor approximation $\sqrt{1+x} \approx 1 + \frac{x}{2} - \frac{x^2}{8}$ \cite{primer_on_nf_bf, mw_nf_mimo_resolution}, yielding
\begin{align}
    \label{eq:dist_approx}
    d_m^{\mathrm{tx}} &\approx d - m d_{\mathrm{a}} \sin{(\theta)} + \frac{(m d_{\mathrm{a}})^2 \cos^2{(\theta)}}{2d}.
\end{align}

\begin{figure}[tb]
    \centering
    \includegraphics[width=\linewidth]{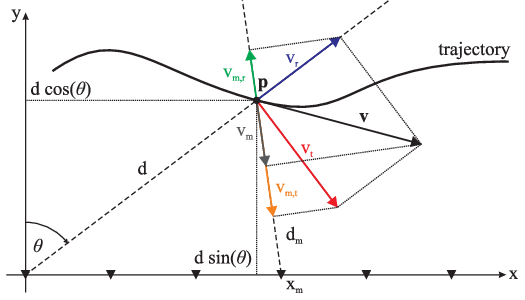}
    \caption{Considered geometry and velocity vectors.}
    \label{fig:ula_geom_with_vel}
\end{figure}

Consider that the point of interest $\mathbf{p}$ is moving with velocity vector expressed in polar coordinates by its radial and tangential components $\mathbf{v} = [v_{\mathrm{r}}, v_{\mathrm{t}}] ^T$.
In the array near-field, due to the proximity of the target and the extent of the array, each individual antenna observes a different range rate.
A single antenna, corresponding to a SISO system, is only able to observe radial velocity with respect to its own position.
The range rate observed by a single antenna is equal to the projection of the velocity vector $\mathbf{v}$ along the direction of the baseline connecting the antenna and the point of interest $\mathbf{p}$.
The geometry and the considered velocity vectors are illustrated in Fig. \ref{fig:ula_geom_with_vel}.
The radial and tangential components of the velocity vector at a point $\mathbf{p}$ for the $m$th antenna are calculated by projections as follows \cite{nf_dopp_perf_bounds, nf_dopp_pred_beamform}
\begin{align}
    \label{eq:mth_ant_vel_proj}
    v_{m, \mathrm{r}}^{\mathrm{tx}}(\mathbf{p}, \mathbf{v}) &= \frac{d - m d_{\mathrm{a}}\sin{(\theta)}}{d_m^{\mathrm{tx}}} v_{\mathrm{r}}\\
    v_{m, \mathrm{t}}^{\mathrm{tx}}(\mathbf{p}, \mathbf{v}) &= \frac{m d_{\mathrm{a}} \cos{(\theta)}}{d_m^{\mathrm{tx}}} v_{\mathrm{t}}.
\end{align}
Expanding with the per-antenna distance approximation from \eqref{eq:dist_approx} gives
\begin{align}
    \label{eq:rad_vel_expanded}
    v_{m, \mathrm{r}}^{\mathrm{tx}}(\mathbf{p}, \mathbf{v}) &= \frac{d - m d_{\mathrm{a}}\sin{(\theta)}}
    {d - m d_{\mathrm{a}} \sin{(\theta)} + \frac{(m d_{\mathrm{a}})^2 \cos^2{(\theta)}}{2d}} v_{\mathrm{r}}\\
    \label{eq:tan_vel_expanded}
    v_{m, \mathrm{t}}^{\mathrm{tx}}(\mathbf{p}, \mathbf{v}) &= \frac{m d_{\mathrm{a}} \cos{(\theta)}}
    {d - m d_{\mathrm{a}} \sin{(\theta)} + \frac{(m d_{\mathrm{a}})^2 \cos^2{(\theta)}}{2d}} v_{\mathrm{t}}.
\end{align}
The range rate observed by the $m$th antenna is a sum of the projections of the radial and tangential velocity components as follows
\begin{align}
    \label{eq:mth_ant_vel}
    v_{m}^{\mathrm{tx}}(\mathbf{p}, \mathbf{v}) =  v_{m,r}^{\mathrm{tx}}(\mathbf{p}, \mathbf{v}) + v_{m, \mathrm{t}}^{\mathrm{tx}}(\mathbf{p}, \mathbf{v}).
\end{align}
The bistatic velocity is given by $ v_{m, n}(\mathbf{p}, \mathbf{v}) =  v_{m}^{\mathrm{tx}}(\mathbf{p}, \mathbf{v}) +  v_{n}^{\mathrm{rx}}(\mathbf{p}, \mathbf{v})$.

\subsection{Propagation channel model}

Given that the point of interest $\mathbf{p}$ is located within the near-field of the aperture, the propagation channel is modeled as a line-of-sight. 
In the considered array near-field region given by $[1.2D, d_{\mathrm{FA}}/6.952]$, the amplitude differences across the array due to path loss can be considered negligible, and the LOS channel simplifies to a phase shift \cite{primer_on_nf_bf}.
The propagation channel has two components: the first is the phase shift due to the position of point $\mathbf{p}$, and the second is the phase shift due to motion. Assuming constant velocity $\mathbf{v}$ over the processing interval, the bistatic channel can be written as
\begin{align}
    \label{eq:chan}
    h_{m, n}(\mathbf{p}, f, \mathbf{v}, t) &= e^{-j2\pi \frac{f}{c} d_{m, n}(\mathbf{p})} e^{-j2\pi\frac{f}{c} v_{m, n}(\mathbf{p}, \mathbf{v}) t }.
    % &= h_{\mathrm{R},m}(\mathbf{p}, f) h_{m, \mathrm{v}}(\mathbf{p}, f, \mathbf{v}, t)
\end{align}
where $f$ is the operating frequency and $c$ is the speed of light. 
Consider that the system is narrowband and operates at a carrier frequency $f_{\mathrm{c}}$. The transmitted or received signal bandwidth $B$ is much smaller than the carrier frequency, $B \ll f_{\mathrm{c}}$. 
This allows to replace the frequency variable $f$ with a constant $f_{\mathrm{c}}$ in \eqref{eq:chan} yielding
\begin{align}
    \label{eq:nb_chan}
    h_{m, n}(\mathbf{p}, f_{\mathrm{c}}, \mathbf{v}, t) &= e^{-j2\pi \frac{f_{\mathrm{c}}}{c} d_{m, n}(\mathbf{p})} e^{-j2\pi\frac{f_{\mathrm{c}}}{c} v_{m, n}(\mathbf{p}, \mathbf{v}) t } \nonumber \\
    &= h_{m,n}^{\mathrm{d}}(\mathbf{p}, f_{\mathrm{c}}) 
    h_{m,n}^{\mathrm{v}}(\mathbf{p}, f_{\mathrm{c}}, \mathbf{v}, t),
\end{align}
where $h_{m,n}^{\mathrm{d}}$ denotes the channel component due to static distance and $h_{m,n}^{\mathrm{v}}$ the channel component due to velocity.
Under the narrowband assumption, the frequency shift due to Doppler can be considered constant across frequency. The target is assumed to be an isotropic scatterer with the same complex response across all observation angles and antennas.

\subsection{Ambiguity function}
The ambiguity function definition follows that of a matched filter. The matched filter response is obtained by multiplying the observed channel by the complex conjugate of the hypothetical channel response and summing across the antenna elements and integrating over the observation time. Because the analysis considers a narrowband channel evaluated at the carrier frequency, no integration over frequency is included. The result is a spatial-velocity ambiguity function rather than the delay-Doppler ambiguity function, which is dependent on a transmitted waveform.
Considering a target at $\mathbf{p}' = [d', \theta']^T$ with velocity $\mathbf{v}' = [v'_{\mathrm{r}}, v'_{\mathrm{t}}]^T$, the ambiguity function is defined as
\begin{align}
    \label{eq:amb_func_nb}
    \mathcal{A}(\mathbf{p}', \mathbf{p}, \mathbf{v}', \mathbf{v}) 
    &= \frac{1}{\sqrt{MN}} \sum_{m \in \mathcal{M}} \sum_{n \in \mathcal{N}}
    \int_{-\infty}^{+\infty}  
    h_{m, n}(\mathbf{p}', f_{\mathrm{c}}, \mathbf{v}', t) \nonumber \\
    & \qquad\qquad\qquad \times
    h_{m,n}^*(\mathbf{p}, f_{\mathrm{c}}, \mathbf{v}, t)
    \,dt.
\end{align}
Consider an integration period $T$ over which the ambiguity function is calculated. 
In practice, due to sampling, the integration period would be implemented as a summation over the slow time samples. Here, the number of received slow-time samples is assumed to be large so that the integration can be accurately approximated as a continuous function. The continuous formulation facilitates the derivation of closed-form expressions of the ambiguity function and velocity resolution.
Separating the channel into the range and Doppler components from \eqref{eq:nb_chan} and updating the integration bounds gives
\begin{align}
    \label{eq:amb_func_nb_expanded}
    &\mathcal{A}(\mathbf{p}', \mathbf{p}, \mathbf{v}', \mathbf{v}) 
    = \frac{1}{\sqrt{MN}} \sum_{m \in \mathcal{M}} \sum_{n \in \mathcal{N}}
    h_{m,n}^{\mathrm{d}}(\mathbf{p}', f_{\mathrm{c}})
    h_{m,n}^{\mathrm{d}*}(\mathbf{p}, f_{\mathrm{c}}) \nonumber \\
    & \qquad  \times\frac{1}{T}\int_{0}^{T}
    h_{m,n}^{\mathrm{v}}(\mathbf{p}', f_{\mathrm{c}}, \mathbf{v}', t)
    h_{m,n}^{\mathrm{v} *} (\mathbf{p}, f_{\mathrm{c}}, \mathbf{v}, t)
    \,dt.
\end{align}
To investigate the Doppler ambiguity function, consider a cross-cut at the target position $\mathbf{p} = \mathbf{p}'$. The product of the channel component due to static distance simplifies to $h_{m,n}^{\mathrm{d}}(\mathbf{p}', f_{\mathrm{c}}) h_{m,n}^{\mathrm{d}*}(\mathbf{p}, f_{\mathrm{c}})  = 1$, yielding
\begin{align}
    \label{eq:amb_func_dopp}
    &\mathcal{A}(\mathbf{p}', \mathbf{v}', \mathbf{v}) = \\
    &= \frac{1}{\sqrt{MN}} \sum_{m \in \mathcal{M}} \sum_{n \in \mathcal{N}}
    \frac{1}{T}\int_{0}^{T}
    e^{-j 2 \pi \frac{f_{\mathrm{c}}}{c} \Delta v_{m,n} (\mathbf{p}', \mathbf{v}', \mathbf{v}) t}
    \,dt, \nonumber
    % \\&= \frac{1}{\sqrt{MN}} \sum_{m \in \mathcal{M}} \sum_{n \in \mathcal{N}}
    % e^{-j \pi \frac{f_{\mathrm{c}}}{c} \Delta v_{m,n} (\mathbf{p}', \mathbf{v}', \mathbf{v}) T} \sinc{\left( \frac{f_{\mathrm{c}}}{c} \Delta v_{m,n} (\mathbf{p}', \mathbf{v}', \mathbf{v})T \right)}
\end{align}
where $\Delta v_{m,n} (\mathbf{p}', \mathbf{v}', \mathbf{v}) = v_{m,n} (\mathbf{p}', \mathbf{v}') - v_{m,n} (\mathbf{p}', \mathbf{v})$ is the bistatic velocity difference observed at target position $\mathbf{p}'$.
To obtain closed-form expressions, the velocity ambiguity function is separated into radial and tangential components, which are analyzed separately.

\subsection{Radial velocity ambiguity function}

Expanding with \eqref{eq:rad_vel_expanded}, the radial velocity difference can be written as 
\begin{align}
    \label{eq:radial_vel_diff}
    &\Delta v_{m, \mathrm{r}}^{\mathrm{tx}} (\mathbf{p}', \mathbf{v}', \mathbf{v}) 
    = v_{m, \mathrm{r}}^{\mathrm{tx}} (\mathbf{p}', \mathbf{v}') - v_{m, \mathrm{r}}^{\mathrm{tx}} (\mathbf{p}', \mathbf{v}) \nonumber \\
    % &=\left(  \frac{d' - m d_{\mathrm{a}}\sin{(\theta')}}
    % {d_m(\mathbf{p}')} v'_{\mathrm{r}}
    % - \frac{d' - m d_{\mathrm{a}}\sin{(\theta')}}
    % {d_m(\mathbf{p}')} v_{\mathrm{r}}
    % \right) \\
    % &= \frac{d' - m d_{\mathrm{a}}\sin{(\theta')}}
    % {d_m(\mathbf{p}')} 
    % (v'_{\mathrm{r}} - v_{\mathrm{r}}) \\
    % &\qquad =\frac{d' - m d_{\mathrm{a}}\sin{(\theta')}}
    % {d' - m d_{\mathrm{a}} \sin{(\theta')} + \frac{(m d_{\mathrm{a}})^2 \cos^2{(\theta')}}{2d'}} 
    % (v'_{\mathrm{r}} - v_{\mathrm{r}}) \nonumber \\
    &\qquad = 
    \frac{1}
    {1 + \frac{(m d_{\mathrm{a}})^2 \cos^2{(\theta')}}{2d'(d' - m d_{\mathrm{a}} \sin{(\theta')})}}
    (v'_{\mathrm{r}} - v_{\mathrm{r}}).
\end{align}
To arrive at a closed-form expression, the complex exponent term is approximated.
In the considered near-field region, the maximum value of the term $\frac{(m d_{\mathrm{a}})^2 \cos^2{(\theta')}}{2d'(d' - m d_{\mathrm{a}}\sin{(\theta')})} $ is less than $ 0.1$ across all $m$ for $d' \geq 1.2D$.
This approximation neglects the impact of the aperture on the radial velocity and coincides with the velocity difference observed in a single antenna (SISO) configuration 
\begin{align}
    \label{eq:rad_vel_diff_approx}
    \Delta v_{m, \mathrm{r}}^{\mathrm{tx}} (\mathbf{p}', \mathbf{v}', \mathbf{v})  \approx v'_{\mathrm{r}} - v_{\mathrm{r}}.
\end{align}

\subsubsection{SIMO/MISO}
Consider that only the transmit or receive aperture is large while the other consists of a single antenna element. Here, the derivation is done with respect to a large transmit aperture and a single receive element by setting $N=1$. By reciprocity, the same results apply to a large receive aperture with a single transmitter.
Calculating the bistatic velocity difference with \eqref{eq:rad_vel_diff_approx} and substituting into \eqref{eq:amb_func_dopp} and solving the integral, the radial velocity ambiguity function can be approximated as
\begin{align}
    \label{eq:rad_vel_amb_approx}
    &\mathcal{A}_{\mathrm{SIMO/MISO}}(d', v'_{\mathrm{r}}, v_{\mathrm{r}})
    % &\approx
    % \frac{1}{T}\int_{0}^{T} e^{-j 2 \pi \frac{f_{\mathrm{c}}}{c}  2(v'_{\mathrm{r}} - v_{\mathrm{r}}) t }
    % \cancelto{1}{\frac{1}{D} \sqrt{\frac{2 \pi}{\alpha}} \left( C(X) + j S(X) \right)}\, dt \\
    \approx
    \frac{\sqrt{M}}{T}\int_{0}^{T} e^{-j 4 \pi \frac{f_{\mathrm{c}}}{c} (v'_{\mathrm{r}} - v_{\mathrm{r}}) t }\, dt \nonumber \\
    & \qquad\qquad = \sqrt{M} e^{-j \pi  \frac{2 (v'_{\mathrm{r}} - v_{\mathrm{r}})}{\lambda} T }\sinc{\left( \frac{2 (v'_{\mathrm{r}} - v_{\mathrm{r}})}{\lambda} T \right)}.
\end{align}

\begin{figure}[tb]
    \centering
    \includegraphics[width=\linewidth]{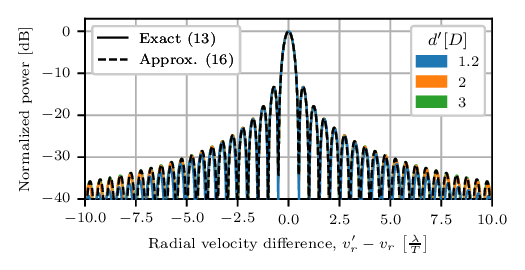}
    \caption{Radial velocity ambiguity function of ULA in SIMO/MISO configuration, $D= 50 \lambda$, $d_{\mathrm{a}} = 0.5 \lambda$ and selected target distances at boresight $\theta' = 0 \degree$.}
    \label{fig:vr_cc_simo_miso}
\end{figure}

Fig. \ref{fig:vr_cc_simo_miso} illustrates the comparison between the exact and approximated radial velocity ambiguity function. The largest discrepancy can be observed in the sidelobes for the target at minimum distance from the array. Note that the approximation captures well the mainlobe width,  which determines the resolution.
The velocity resolution is evaluated in terms of the half-power ($-3$ dB) beamwidth of the ambiguity function.
The resolution is obtained by numerically solving for the argument for which the power of the ambiguity function reaches half of the maximum value.
Solving for the radial resolution of the SIMO/MISO configuration yields
\begin{align}
    \label{eq:vr_res_simo_miso}
    % \sinc^2{\left( \frac{2 (v'_{\mathrm{r}} - v_{\mathrm{r}})}{\lambda} T \right)} &= 0.5 \\
    % \sinc{\left( x_{3 \mathrm{dB}} \right)} &= \frac{1}{\sqrt{2}} \\
    % x_{3 \mathrm{dB}} &= 0.443 \\
    % \frac{2 (v'_{\mathrm{r}} - v_{\mathrm{r}})}{\lambda} T &= 2 \cdot 0.443  \\
    \Delta v_{\mathrm{r, SIMO/MISO}} &= 0.443 \frac{\lambda}{T}.
\end{align}

\begin{figure}[tb]
    \centering
    \includegraphics[width=\linewidth]{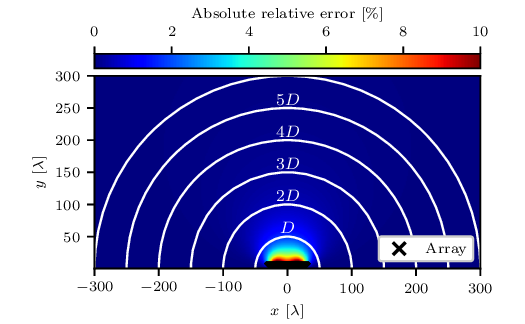}
    \caption{Distribution of absolute relative radial velocity resolution error in space for SIMO/MISO configuration, $D= 50 \lambda$ and $d_{\mathrm{a}} = 0.5 \lambda$.}
    \label{fig:vr_err_map_simo_miso}
\end{figure}

Fig. \ref{fig:vr_err_map_simo_miso} shows the relative resolution error between the exact and approximated radial velocity ambiguity function. 
Despite neglecting the nonlinear term in \eqref{eq:radial_vel_diff}, the approximation for SIMO/MISO configuration remains accurate with relative error below $1.3\%$ across the considered near-field region. 
When considering radial velocity, the near-field region does not offer any significant improvement in terms of resolution; only slightly improved sidelobe levels decay, as shown in Fig. \ref{fig:vr_cc_simo_miso}. A similar conclusion can be reached by investigating the CRLB in \cite{nf_dopp_perf_bounds}.

\subsubsection{MIMO}
Since the radial velocity difference in \eqref{eq:rad_vel_diff_approx} is independent of the antenna index, the resulting approximation of the radial velocity ambiguity function remains unchanged across different array configurations, including SIMO/MISO, MIMO and a collection of SISO (CSISO) radars. Considering different array configurations in the context of radial velocity does not improve the ambiguity function, resolution or the sidelobe level (SL).
The only difference between the array configurations is the array processing gain, which depends on the number of antenna elements.
However, a slight reduction in approximation accuracy is expected in MIMO and CSISO configurations due to doubling of the velocity approximation error for monostatic paths.

\subsection{Tangential velocity ambiguity function}

The tangential velocity difference is obtained from \eqref{eq:tan_vel_expanded} and can be written as
\begin{align}
    \label{eq:tan_vel_diff_expanded}
    &\Delta v_{m, \mathrm{t}}^{\mathrm{tx}} (\mathbf{p}', \mathbf{v}', \mathbf{v}) 
    = v_{m, \mathrm{t}}^{\mathrm{tx}} (\mathbf{p}', \mathbf{v}') - v_{m, \mathrm{t}}^{\mathrm{tx}} (\mathbf{p}', \mathbf{v}) \nonumber \\
    % &= \frac{m d_{\mathrm{a}} \cos{(\theta')}}{d_m(\mathbf{p}')} v'_{\mathrm{t}}
    % - \frac{m d_{\mathrm{a}} \cos{(\theta')}}{d_m(\mathbf{p}')} v_{\mathrm{t}}  \\
    % &= \frac{m d_{\mathrm{a}} \cos{(\theta')}}{d_m(\mathbf{p}')} 
    % (v'_{\mathrm{t}} - v_{\mathrm{t}}) \\
    &\qquad= \frac{m d_{\mathrm{a}} \cos{(\theta')}}
    {d' - m d_{\mathrm{a}} \sin{(\theta')} + \frac{(m d_{\mathrm{a}})^2 \cos^2{(\theta')}}{2d'}} 
    (v'_{\mathrm{t}} - v_{\mathrm{t}}).
\end{align}
Similarly, as before, to arrive at a closed-form expression of the velocity ambiguity function, the velocity difference in \eqref{eq:tan_vel_diff_expanded} needs to be approximated. Here, the first-order approximation is considered, which neglects the antenna-dependent terms in the denominator, yielding
\begin{align}
    \label{eq:tan_vel_diff_approx}
    &\Delta v_{m, \mathrm{t}}^{\mathrm{tx}} (\mathbf{p}', \mathbf{v}', \mathbf{v}) \approx \frac{ m d_{\mathrm{a}} \cos{(\theta')}}{d'} (v'_{\mathrm{t}} - v_{\mathrm{t}}).
\end{align}
This approximation is equivalent to the far-field approximation $(d' \gg x)$ as the wavefront curvature across the array is considered negligible. Despite ignoring the quadratic phase term characteristic of the near-field, the resulting ambiguity function approximation remains accurate.
Note that the tangential velocity difference in \eqref{eq:tan_vel_diff_approx} is a function of the antenna index. Consequently, the resulting ambiguity function will be a function of antenna configuration, the total array aperture and the distance from the array.

\subsubsection{SIMO/MISO}
Consider that only the transmit or receive aperture is large while the other consists of a single antenna element.
To obtain a closed-form expression, the discrete sum over the antenna elements is approximated as an integral based on the continuous aperture approximation \cite{primer_on_nf_bf}. This approximation is valid when the number of antennas is large and the element spacing is dense and satisfies the Nyquist criterion. In the considered system, the antennas are distributed symmetrically around the origin and implement an aperture $D = (M-1) d_{\mathrm{a}}$. Substituting \eqref{eq:tan_vel_diff_approx} into \eqref{eq:amb_func_dopp} and approximating the sum as an integral over the aperture yields
% \begin{align}
%     \mathcal{A}(d', v'_{\mathrm{t}}, v_{\mathrm{t}})
%     &= \frac{1}{T}\int_{0}^{T} \frac{1}{\sqrt{M}} \sum_{(-(M-1)/2}^{(M-1)/2}
%     e^{-j2 \pi \frac{f_{\mathrm{c}}}{c} 
%     \frac{ m d_{\mathrm{a}} \cos{(\theta')}}{d'}
%     (v'_{\mathrm{t}} - v_{\mathrm{t}}) t
%     } \,dt \\
%     &= \frac{1}{T}\int_{0}^{T} \frac{\sqrt{M}}{D} 
%     \int_{-(M-1)/2}^{(M-1)/2}
%     e^{-j2 \pi \frac{f_{\mathrm{c}}}{c} 
%     \frac{ m d_{\mathrm{a}} \cos{(\theta')}}{d'}
%     (v'_{\mathrm{t}} - v_{\mathrm{t}}) t
%     } dm\, dt. \nonumber
% \end{align}
\begin{align}
    \label{eq:tan_vel_amb_int_approx}
    &\mathcal{A}_{\mathrm{SIMO/MISO}}(d', v'_{\mathrm{t}}, v_{\mathrm{t}}) \approx  \\ 
    &\qquad \approx\frac{1}{T} \int_{0}^{T} 
    \frac{\sqrt{M}}{D} \int_{-D/2}^{D/2}    
    e^{-j2 \pi \frac{f_{\mathrm{c}}}
    {c} \frac{x \cos{(\theta')}}{d'} 
    (v'_{\mathrm{t}} - v_{\mathrm{t}}) t} \, dx \, dt . \nonumber
\end{align}
The integral over the aperture in \eqref{eq:tan_vel_amb_int_approx} can be expressed as
\begin{align}
    \label{eq:dap_int}
    \frac{1}{D} \int_{-D/2}^{D/2}    
    e^ {-j2 \pi \frac{f_{\mathrm{c}}}{c} \frac{x \cos{(\theta')}}{d'} (v'_{\mathrm{t}} - v_{\mathrm{t}}) t} \, dx 
    % &= \frac{\sin{( \pi \beta D t) }}{ \pi \beta D t} \nonumber \\
    &= \sinc{(\beta D t)},
\end{align}
where $\beta =  \frac{(v'_{\mathrm{t}} - v_{\mathrm{t}})}{d' \lambda} \cos{(\theta')}$.
Finally, substituting \eqref{eq:dap_int} into \eqref{eq:tan_vel_amb_int_approx} and calculating the integral with respect to time yields
\begin{align}
    \label{eq:tan_vel_approx_simo_miso}
    \mathcal{A}_{\mathrm{SIMO/MISO}}(d', v'_{\mathrm{t}}, v_{\mathrm{t}}) 
    &= \frac{\sqrt{M}}{T}\int_{0}^{T} \sinc{(\beta D t)} \, dt \nonumber \\
    % &= \frac{d' \lambda}{\cos{(\theta')}} \frac{\sqrt{M}}{ (v'_{\mathrm{t}} - v_{\mathrm{t}})T} \Si{\left( (v'_{\mathrm{t}} - v_{\mathrm{t}})  \frac{D\cos{(\theta')}}{d' \lambda} T\right)} \\
    &= \frac{\sqrt{M}}{\beta D T} \Si{(\beta D T)},
\end{align}
where $\Si{(x) = \int_{0}^{x} \sinc{(x)} \, dx} = \int_{0}^{x} \sin{(\pi x)} / (\pi x) \, dx$ is the normalized sine integral function.
Given the ambiguity function, the tangential resolution for the SIMO/MISO configuration is
\begin{align}
    \label{eq:vt_res_simo_miso}
    % \left(\frac{\Si{(\beta D T)}}{\beta D T} \right)^2 &= 0.5 \\
    % \frac{\Si{(x_{3 \mathrm{dB}})}}{x_{3 \mathrm{dB}}} &= \frac{1}{\sqrt{2}} \\
    % x_{3 \mathrm{dB}} &= 0.803 \\
    % \frac{(v'_{\mathrm{t}} - v_{\mathrm{t}})}{d' \lambda} \cos{(\theta')} DT &= 2 \cdot 0.803 \\
    \Delta v_{\mathrm{t, SIMO/MISO}} &= 1.606 \frac{d'}{D \cos{(\theta ')}} \frac{\lambda}{T}.
\end{align}

\begin{figure}[tb]
    \centering
    \includegraphics[width=\linewidth]{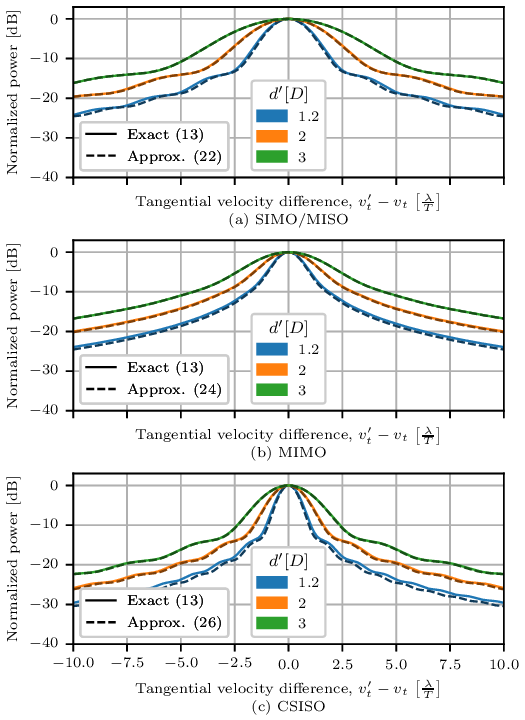}
    \caption{Tangential velocity ambiguity function for different array configurations, $D= 50 \lambda$, $d_{\mathrm{a}} = 0.5 \lambda$ and selected target distances at boresight $\theta' = 0 \degree$.}
    \label{fig:vt_cc_combined}
\end{figure}

\begin{figure}[tb]
    \centering
    \includegraphics[width=\linewidth]{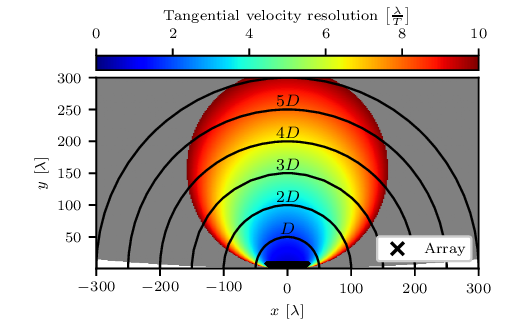}
    \caption{Tangential velocity resolution as a function of position for SIMO/MISO configuration, $D= 50 \lambda$ and $d_{\mathrm{a}} = 0.5 \lambda$.}
    \label{fig:vt_res_map_simo_miso}
\end{figure}

\begin{figure}[tb]
    \centering
    \includegraphics[width=\linewidth]{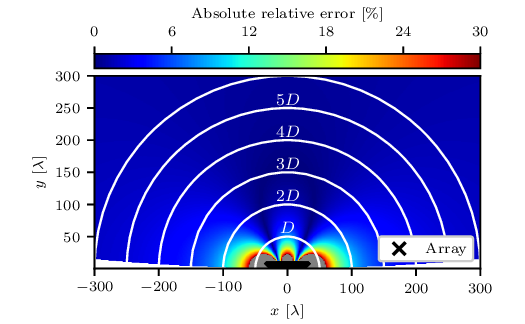}
    \caption{Distribution of absolute relative tangential velocity resolution error in space for SIMO/MISO configuration, $D= 50 \lambda$ and $d_{\mathrm{a}} = 0.5 \lambda$.}
    \label{fig:vt_err_map_simo_miso}
\end{figure}

Fig. \ref{fig:vt_cc_combined}a illustrates the comparison between the exact and approximated tangential ambiguity function for the SIMO/MISO configuration. 
The tangential velocity resolution depends on the target distance, improving as the target approaches the array. Despite utilizing the far-field approximation in \eqref{eq:tan_vel_diff_approx}, the resulting approximation remains accurate and reproduces well the mainlobe and the sidelobes of the ambiguity function.
Fig. \ref{fig:vt_res_map_simo_miso} shows the map of exact tangential velocity resolution as a function of distance and angle with respect to the array for the SIMO/MISO configuration.
As the argument in \eqref{eq:tan_vel_approx_simo_miso} is a function of $\cos{(\theta')}$, the resolution degrades with increasing target angle.
Fig. \ref{fig:vt_err_map_simo_miso} shows the absolute relative tangential velocity resolution error between the exact and approximated expressions. The largest error of around 30\% appears at large angles and distances below $2D$ from the array.

\subsubsection{MIMO}
Since the tangential velocity approximation in \eqref{eq:tan_vel_diff_approx} is a function of antenna index, different array configurations will yield different ambiguity functions. Consider identical collocated large transmit and receive antenna arrays $(M=N)$. 
As the transmit and receive antenna elements are collocated, no virtual array aperture extension is exploited and the total virtual aperture is the same as the physical aperture.
The double summation over the transmit and receive antenna elements in \eqref{eq:amb_func_dopp} simplifies to a square of a single sum, which corresponds to the squared aperture integral from \eqref{eq:dap_int}.
The ambiguity function of MIMO configuration with identical and collocated TX and RX apertures can be written as
\begin{align}
    \label{eq:tan_vel_approx_mimo}
    &\mathcal{A}_{\mathrm{MIMO}}(d', v'_{\mathrm{t}}, v_{\mathrm{t}}) 
    = \frac{\sqrt{MN}}{T}\int_{0}^{T} \sinc^2{(\beta D t)} \, dt \\
    &\qquad\quad = \frac{\sqrt{MN}}{\beta D T} \left( \Si{(2\beta D T)} + \frac{\cos{\left( 2 \pi \beta D T \right) - 1 }}{2 \pi^2 \beta D T}\right). \nonumber
\end{align}
Given the ambiguity function, the tangential resolution for the MIMO configuration is
\begin{align}
    \label{eq:vt_res_mimo}
    % \left(\frac{\Si{(2\beta D T)}}{\beta D T} + \frac{\cos{\left( 2 \pi \beta D T \right) - 1 }}{2( \pi \beta D T)^2}\right)^2 &= 0.5 \\
    % \frac{\Si{(2\beta D T)}}{\beta D T} 
    % + \frac{\cos{\left( 2 \pi \beta D T \right) - 1 }}{2( \pi \beta D T)^2} &= \frac{1}{\sqrt{2}} \\
    % \frac{\Si{(2x_{3 \mathrm{dB}})}}{x_{3 \mathrm{dB}}} 
    % + \frac{\cos{\left( 2 \pi x_{3 \mathrm{dB}} \right) - 1 }}{2 \pi x_{3 \mathrm{dB}}^2} &= \frac{1}{\sqrt{2}} \\
    % x_{3 \mathrm{dB}} &= 0.59 \\
    % \frac{(v'_{\mathrm{t}} - v_{\mathrm{t}})}{d' \lambda} \cos{(\theta')} DT &= 2 \cdot 0.59 \\
    \Delta v_\mathrm{{t, MIMO}} &= 1.179 \frac{d'}{D \cos{(\theta ')}} \frac{\lambda}{T}.
\end{align}

Fig. \ref{fig:vt_cc_combined}b illustrates the tangential velocity ambiguity function for the MIMO configuration with identical TX and RX apertures. Due to the squared aperture integral, the MIMO configuration offers improved resolution; however, the sidelobe level remains mostly the same.

\subsubsection{Collection of SISO (CSISO) radars}

Consider identical large collocated TX and RX apertures $(M=N)$, where each receiver only processes the signal transmitted by its corresponding collocated transmitter. Since no multiplexing scheme exploiting the bistatic path across different transmitters and receivers is employed, the array can be interpreted as a collection of SISO radars (CSISO).
For the CSISO configuration, the double sum in \eqref{eq:amb_func_dopp} simplifies to a single sum over the antenna elements with the velocity difference from \eqref{eq:tan_vel_diff_approx} doubled in the exponential due to per-element monostatic propagation. 
Doubling the exponent argument and approximating the sum over the antenna elements as an integral yields
\begin{align}
    \label{eq:tan_vel_approx_csiso}
    \mathcal{A}_{\mathrm{CSISO}}(d', v'_{\mathrm{t}}, v_{\mathrm{t}}) 
    &= \frac{\sqrt{M}}{T}\int_{0}^{T} \sinc{(2 \beta D t)} \, dt \nonumber \\
    % &= \frac{d' \lambda}{2\cos{(\theta')}} \frac{\sqrt{M}}{ (v'_{\mathrm{t}} - v_{\mathrm{t}})T} \Si{\left(2 (v'_{\mathrm{t}} - v_{\mathrm{t}})  \frac{D\cos{(\theta')}}{d' \lambda} T\right)} \\
    &= \frac{\sqrt{M}}{2\beta D T} \Si{(2 \beta D T)}.
\end{align}
Given the ambiguity function, the tangential resolution of the CSISO configuration is
\begin{align}
    % \left( \frac{ \Si{(2 \beta D T)} }{2\beta D T} \right)^2 &= 0.5 \\
    % \frac{ \Si{(2 \beta D T)} }{2\beta D T} &= \frac{1}{\sqrt{2}} \\
    % \frac{ \Si{(2  x_{3 \mathrm{dB}})} }{2  x_{3 \mathrm{dB}}} &= \frac{1}{\sqrt{2}} \\
    % x_{3 \mathrm{dB}} &= 2 \cdot 0.402 \\
    \Delta v_\mathrm{{t, CSISO}} &= 0.803 \frac{d'}{D\cos{(\theta ')}} \frac{\lambda}{T}.
\end{align}
Due to per-element monostatic propagation, the CSISO configuration offers a factor of two improved tangential velocity resolution as compared to the SIMO/MISO from \eqref{eq:vt_res_simo_miso}.
Fig. \ref{fig:vt_cc_combined}c illustrates the comparison between the exact and approximated ambiguity function of the CSISO configuration. The double argument offers improved resolution and improved decay of the sidelobes of the ambiguity function.

\begin{figure}[b]
    \centering
    \includegraphics[width=\linewidth]{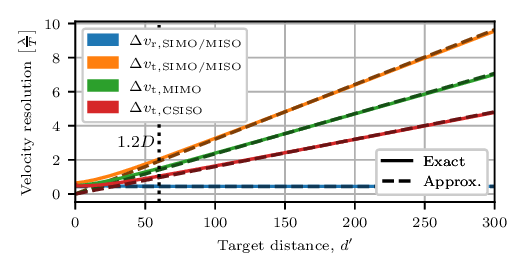}
    \caption{Velocity resolution as a function of target distance $d'$ for target at the boresight $\theta'= 0 \degree$, $D= 50 \lambda$, $d_{\mathrm{a}} = 0.5 \lambda$ and different array configurations.}
    \label{fig:vres_vs_config_vs_tgt_d}
\end{figure}

Fig. \ref{fig:vres_vs_config_vs_tgt_d} illustrates the velocity resolution achieved by the considered array configurations as a function of the target distance. The radial resolution is independent of the target distance and provides the highest resolution. The tangential velocity degrades linearly with target distance. The MIMO configuration offers approximately a $1.36$-fold improvement in the tangential resolution over the SIMO/MISO configuration, while the CSISO achieves the best tangential resolution. Note that the achievable tangential velocity resolution is at least several times poorer than the radial resolution.

\section{Conclusion}
This work investigates the velocity ambiguity function in the near-field of the ULA. 
Closed-form expressions of the velocity ambiguity functions derived using the far-field approximation are shown to remain highly accurate, indicating that the neglected nonlinear terms have a negligible effect.
The radial velocity ambiguity function is largely unaffected by the near-field regime and yields the same ambiguity function and resolution as in the far-field.
On the contrary, the tangential velocity ambiguity function depends on the array configuration and target position.
Consequently, the main benefit of the near-field operation is the ability to determine the target tangential velocity, which is not possible in a far-field SISO system.
For tangential velocity estimation, the best resolution is achieved by a collection of SISO radars, then the MIMO, and finally the SIMO/MISO configuration. The tangential velocity resolution scales linearly with the target distance and exhibits a cosine dependency on the target angle.

\bibliographystyle{IEEEtran}
\bibliography{IEEEabrv.bib, biblio.bib}

\end{document}